\documentclass[10pt]{iopart}

\usepackage{graphicx}
\usepackage{hyperref}
\usepackage{cite}

\begin{document}
\title[]{Tomographic reconstruction of the runaway distribution function in TCV using multispectral synchrotron images}

\author{
T.A. Wijkamp$^{1,2}$, A. Perek$^{2}$, J. Decker$^{3}$, B. Duval$^{3}$, M. Hoppe$^{4}$, G. Papp$^{5}$, U.A. Sheikh$^{3}$, I.G.J. Classen$^{2}$, R.J.E. Jaspers$^{1}$, the TCV team$^{a}$ and the EUROfusion MST1 team$^{b}$
}

\address{$^1$ Department of Applied Physics, Eindhoven University of Technology, Eindhoven 5600 MB, Netherlands}
\address{$^2$ DIFFER - Dutch Institute for Fundamental Energy Research, De Zaale 20, 5612 AJ Eindhoven, the Netherlands}
\address{$^3$ Ecole Polytechnique F\'{e}d\'{e}rale de Lausanne (EPFL), Swiss Plasma Center (SPC), CH-1015 Lausanne, Switzerland}
\address{$^4$ Department of Physics, Chalmers University of Technology, SE-41296 Gothenburg, Sweden}
\address{$^5$ Max Planck Institute for Plasma Physics, D-85748 Garching, Germany}
\address{$^a$ See author list S. Coda et a., \textit{Nuclear Fusion}, 59: 112014 (2019) (\url{https://doi.org/10.1088/1741-4326/ab25cb})}
\address{$^b$ See author list of B. Labit et al. \textit{Nuclear Fusion}, 59: 086020 (2019) (\url{https://doi.org/10.1088/1741-4326/ab2211})}

\begin{abstract}

Synchrotron radiation observed in a quiescent TCV runaway discharge is studied using filtered camera images targeting three distinct wavelength intervals. Through the tomographic SART procedure the high momentum, high pitch angle part of the spatial and momentum distribution of these relativistic particles is reconstructed. Experimental estimates of the distribution are important for verification and refinement of formation-, decay- and transport-models underlying runaway avoidance and mitigation strategy design. Using a test distribution it is demonstrated that the inversion procedure provides estimates accurate to within a few tens of percent in the region of phase-space contributing most to the synchrotron image. We find that combining images filtered around different parts of the emission spectrum widens the probed part of momentum-space and reduces reconstruction errors. Next, the SART algorithm is used to obtain information on the spatiotemporal runaway momentum distribution in a selected TCV discharge. The momentum distribution is found to relax towards an avalanche-like exponentially decaying profile. Anomalously high pitch angles and a radial profile increasing towards the edge are found for the most strongly emitting particles in the distribution. Pitch angle scattering by toroidal magnetic field ripple is consistent with this picture. An alternative explanation is the presence of high frequency instabilities in combination with the formation of a runaway shell at the edge of the plasma.

\end{abstract}

\maketitle
\ioptwocol


\section{Introduction}

Successful operation of the next generation of tokamaks is threatened by runaway electron (RE) formation during a thermal quench. To deal with the issue disruptions should be avoided, and a disruption mitigation system needs to be in place \cite{Hollmann2015,Lehnen2015}. Design of the latter relies on numerical models founded on a detailed understanding of the runaway and background plasma behaviour \cite{Breizman2019}. Experimental verification of theories on both fronts calls for ways to diagnose both (i) conditions in post-disruptive plasmas and (ii) the momentum- and real-space distribution of the runaway electrons.

Runaways gyrating in the tokamak magnetic field \cite{Hoppe2018_1} emit synchrotron radiation. The emission spectrum and geometry are sensitive to details in the RE distribution function, rendering synchrotron radiation a promising source for passively diagnosing runaways \cite{Breizman2019}. Pioneering observations in the 90's \cite{Finken1990,Jaspers1994} sparked synchrotron studies in a wide range of tokamaks over the last two decades, a comprehensive overview of which is provided by Tinguely et al \cite{Tinguely2018_1}. 

Synchrotron spectra are readily obtained with the use of spectrometers. While this approach has been applied with considerable success \cite{Tinguely2018_2}, it has been demonstrated that multiple distributions can be fit to a single spectrum \cite{Stahl2013}. In an effort to provide additional constraints to the distribution function, filtered 2D visible imaging has been employed on multiple devices \cite{Hoppe2018_1,Tinguely2018_1,Shi2010,Yu2013,Zhou2014,Hoppe2018_2,Hoppe2020,Hoppe2020_2}. The downside of using camera images is the loss of spectral information. An attempt has been made to circumvent part of this deficiency by splitting the RGB components of an image \cite{Shi2010}. This approach however suffers from line radiation pollution, obscuring dimmer parts of the pattern.

In this work a novel method for adding the spectral dimension to 2D camera imaging is presented, relying on recent advances in multispectral imaging systems \cite{Perek2019,Perek2020} which are described in section \ref{sec:ch2}. Simultaneous 2D imaging of synchrotron emission in several separated narrow wavelength bands widens the probed region of momentum- and real-space, allowing for tighter constraints on the spatiotemporal evolution of the runaway momentum-distribution. The experimental method is put to the test during TCV runaway discharges, addressed in section \ref{sec:ch3}.

The additional constraints facilitate a tomographic approach to reconstructing the momentum- and real-space distribution of the runaways from synchrotron images. Linear systems of equations linking the distribution and image are generated through the synthetic diagnostic SOFT (Synchrotron-detecting Orbit Following Toolkit) \cite{Hoppe2018_2}, as will be addressed in section \ref{sec:ch4}. The reconstruction approach is outlined in section \ref{sec:ch5}, and is the first attempt to obtain a RE distribution function from synchrotron images without imposing constraints on the distribution shape. Using an avalanche-like test distribution the performance of the algorithm is assessed. It is found that the distribution is reconstructed with an accuracy up to a few tens of percent in the part of phase-space contributing most to synchrotron image.

In section \ref{sec:ch6} the reconstruction approach is applied to TCV experimental data. The momentum-space profile is found to relax towards an exponentially decaying avalanche-like profile. Anomolously high pitch angles and a radial profile increasing with plasma minor radius are found for the well-constrained part of phase-space. Magnetic field ripple and high frequency instabilities are proposed as underlying pitch angle scattering mechanisms. The main conclusions of the paper are revisited in section \ref{sec:ch7}.


\section{Multispectral imaging systems in runaway experiments}
\label{sec:ch2}

\begin{figure}
\centering
\includegraphics[width=0.4\textwidth]{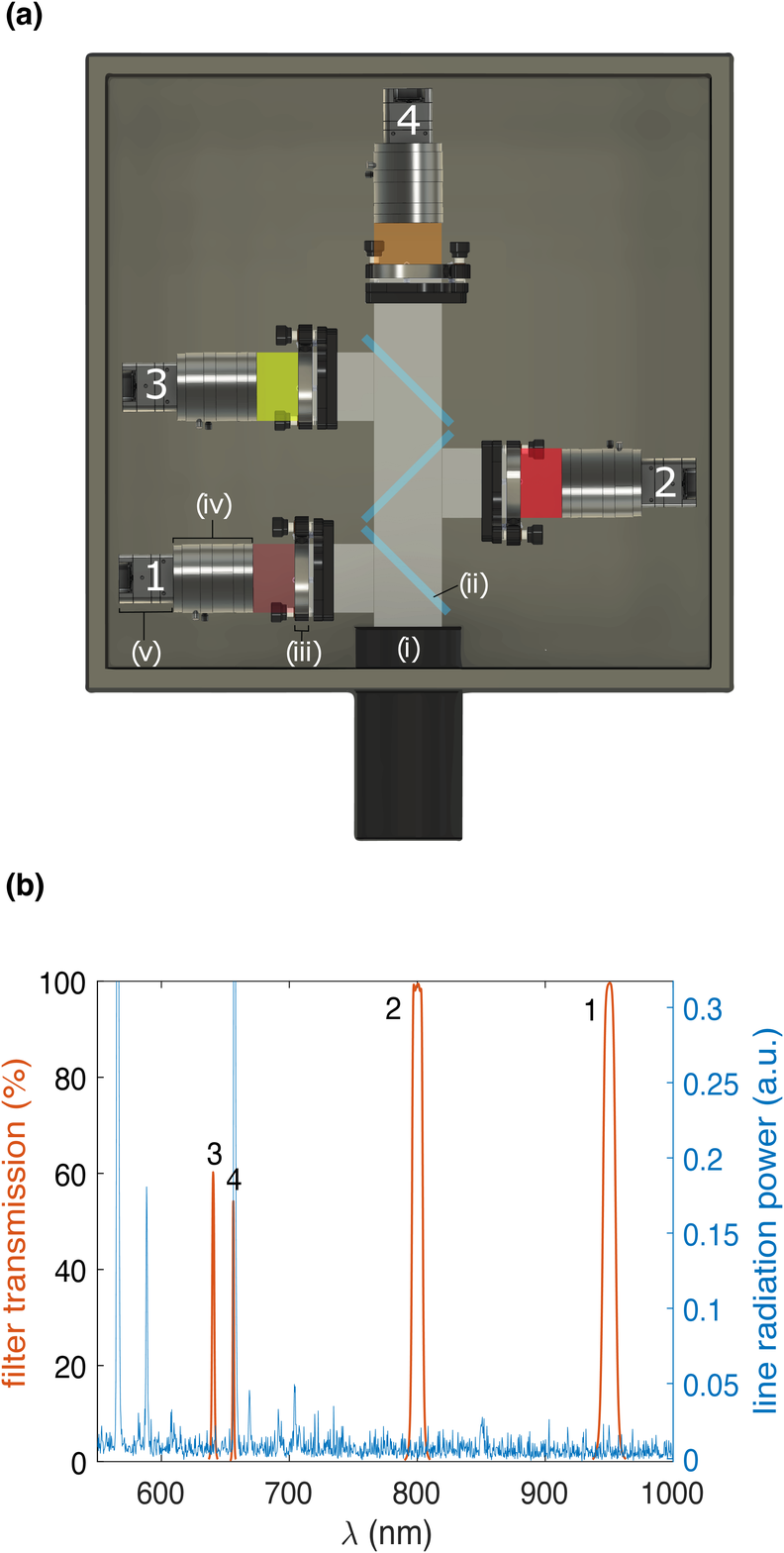}
\caption{Schematic top view of the optical diagnostic MultiCam (a). Light from a tokamak plasma is relayed to the entrance of the box (i). Using a set of beamsplitters (ii) the light is distributed amongst four different camera channels. Each channel has a distinct narrow band optical interference filter (iii), the transmission curves $T$ of which are shown (b) as a function of wavelength $\lambda$. This same plot shows line-radiation recorded by a top-down spectrometer in TCV shot 64717 between 1 and 1.5$\,$s. The transmitted portion of the light is focused on a sensor (v) with the use of an objective (iv).}
\label{fig:1}
\end{figure}

The advances in synchrotron runaway diagnostics described here are made possible by the recent development of multispectral imaging systems in the context of divertor detachment studies \cite{Perek2019,Perek2020}. These imaging devices are capable of recording light from a fusion plasma at several different narrow wavelength intervals using multiple cameras with the same field-of-view. The synchrotron patterns at TCV were gathered using the MultiCam system depicted in figure \ref{fig:1}a. Light enters the system via the relay optic (i), and is subsequently distributed amongst different channels with the use of beam splitters (ii). All channels consist of a filter (iii), objective (iv) and a Ximea camera with Sony IMX252 sensor (v). Hardware and acquisition specifications are identical to that of the optical cavity based multispectral system MANTIS, and are described in detail by Perek et al \cite{Perek2019}. Used camera settings can be found in the supplementary material.

The cameras are equipped with a 950$\,$nm (full-width-half-maximum 10$\,$nm), 800$\,$nm (10$\,$nm), 640.6$\,$nm (1.7$\,$nm) and a deuterium Balmer alpha filter at 656.2$\,$nm (1.1$\,$nm). The former three are placed with the aim of studying synchrotron radiation. The latter is installed to record emission from the cold edge of the plasma. Synchrotron filter transmission ranges were chosen so as to (i) avoid pollution by line emission, (ii) while maximizing the spectral coverage, (iii) keeping in mind the sensitivity curve of the sensors which tends to zero near the infrared. The line-radiation emission spectrum for a typical quiescent runaway shot is shown in figure \ref{fig:1}b, and is obtained by integrating top-down spectrometer data in shot 64717 between 1 and 1.5$\,$s. Note that pollution is low for channels 1-3 as intended. The cameras have the capability of adjusting gain and exposure real time in response to the recorded signal. This means that an optimal signal-to-noise ratio can be achieved within the limits set by the acquisition rate, sensitivity and aperture, whilst avoiding saturation if the intensity increases on a time scale of a few ms or more.

Camera sensors are calibrated with the use of an integrating sphere and a light source with known emission characteristics. Calibrations are performed in full diagnostic configuration with exception of the tokamak optical window, so that optical vignetting losses are accounted for. A signal to photons/s conversion factor is obtained, averaged over the transmission band of the corresponding filter. Calculations show that the inconsistency between the synchrotron spectra and calibration source spectrum for the 10$\,$nm full-width-half-maximum filters leads to relative conversion factor errors which do not exceed 1 percent. Extensive characterization of the readout noise, shot noise and near-saturation non-linear noise has been performed. As a result the signal dependent readout uncertainties are known on a pixel-to-pixel basis.

The position, viewing direction, field-of-view and image distortion of the full optical configuration are obtained after installation of the setup using the EURATOM camera calibration software Calcam \cite{Calcam}. A camera model is fit to a set of matching points in a camera image and detailed TCV CAD model. The outcome is used as input for synthetic diagnostics used in interpreting synchrotron radiation as well as the undistortion of experimentally obtained images. Calibration details are found in the supplementary material.

\begin{figure}
\centering
\includegraphics[width=0.38\textwidth]{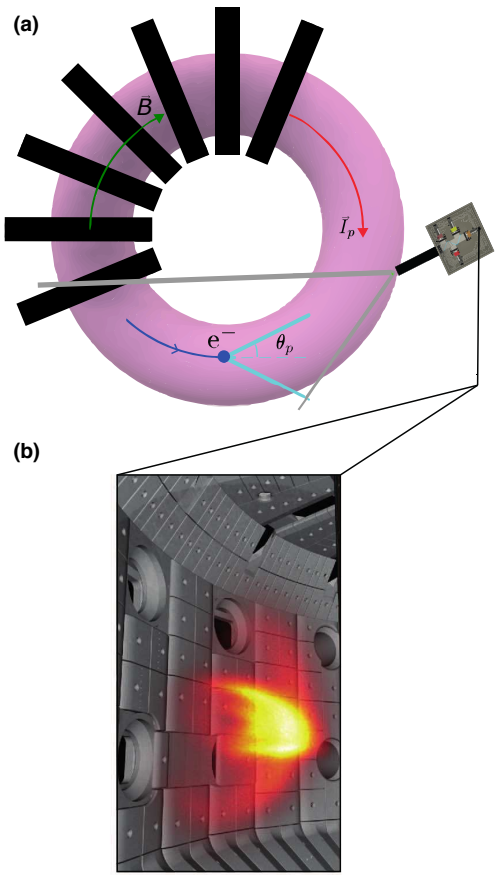}
\caption{MultiCam configuration in the TCV tokamak (a). The MultiCam viewing cone is indicated (gray) along with the magnetic field (green) and plasma current $I_{\mathrm{p}}$ (red) which opposes the runaway movement (blue). The runaways emit radiation along their velocity vector. In a guiding-center picture the relativistic particles follow the magnetic field and emit radiation in a hollow cone along their guiding-center. The half opening angle is equal to the runaway pitch angle $\theta_{\mathrm{p}}$. A typical synchrotron image in TCV is shown (b), overlayed on the calibrated MultiCam view of the TCV CAD model.}
\label{fig:2}
\end{figure}

Positioning of the camera, at central height $z = 0$ m, is in accordance with the counter current runaway movement. As depicted in figure \ref{fig:2}a, the camera has a tangential view of the plasma, facing the REs head-on. This is required as synchrotron emission is strongly concentrated along the RE velocity vector \cite{Hoppe2018_2}. A typical quiescent-phase TCV synchrotron pattern for a runaway beam at camera height is shown in figure \ref{fig:2}b. The undistorted pattern is overlayed on the calibrated view of the TCV CAD model to give an indication of the field-of-view.

\section{Synchrotron radiation in TCV quiescent-phase}
\label{sec:ch3}

Validity assessment of the iterative tomographic reconstruction method presented here is aided by the unique geometry of synchrotron patterns observed in TCV. Synthetic reconstruction of their features requires an accurate estimate of the runaway distribution function. Judging whether or not the estimate is consistent with expectations requires knowledge of the experimental scenario, which is described in this section.

\begin{figure*}
\centering
\includegraphics[width=1\textwidth]{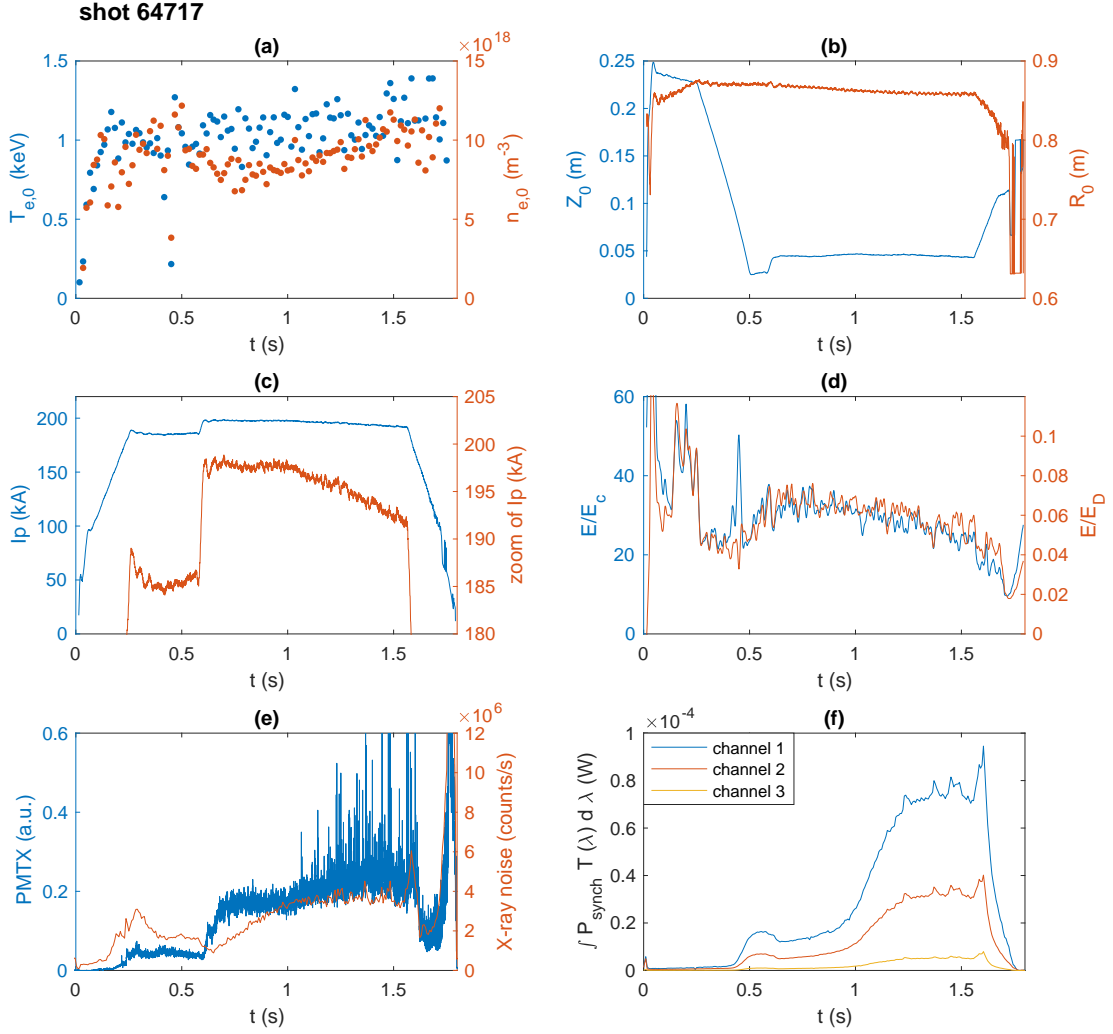}
\caption{Plasma parameter and diagnostic signal evolution during TCV shot 64717. Thomson scattering data provides an estimate for electron temperature and density (a). The plasma density is sufficiently low for the on-axis electric field to exceed the Connor Hastie critical field (d), setting the conditions for runaway production. Significant RE formation is expected on the basis of the ratio of electric field to Dreicer field shown in the same plot. Runaway production is evidenced by a hard X-ray signal (e) from the PMTX and X-ray noise on MulitCam, as well as a rise over time in total recorded intensity on the synchrotron dedicated MultiCam channels (f). During the course of the shot the plasma is translated vertically downwards (b) until 0.65$\,$s, after which a jump in plasma current (c) and PMTX signal is observed.}
\label{fig:3}
\end{figure*}

The synchrotron radiation patterns addressed here appear in the quiescent-phase of a subset of circular plasma TCV runaway discharges with an on axis magnetic field of 1.43 T. Here, shot 64717 is highlighted, whose plasma parameter evolution is shown in figure \ref{fig:3}. Peak plasma density and temperature evolution (a) are based on a radial coordinate mapping of Thomson scattering data \cite{Hawke2017}. By sustaining a sufficiently low electron density and an electric field of 0.2-0.3$\,$V/m, estimated from the temporally smoothed applied loop voltage, the Connor-Hastie critical field \cite{Connor1975} is surpassed (d). Runaway production is evidenced by the bremsstrahlung originating from collisions of runaways with the plasma and wall. Figure \ref{fig:3}e shows the hard X-ray signal from the ex-vessel, uncollimated PhotoMultiplier Tube for hard X-rays (PMTX). This scintillator is positioned 5$\,$m outside the vessel wall, and will only receive X-rays having traversed the intermediate steel components, setting the lower energy bound for detection at roughly $150\,$keV. Another indicator for X-rays is the noise on MultiCam images. The corresponding signal (e) is obtained through the comparison of MultiCam images to their median filtered counterparts.

As opposed to a gas injection triggered disruption scenario, only a small fraction of the current is carried by relativistic electrons. The maximum in the PMTX signal during the flat top of shot 64717 is a factor of 50 lower than that recorded in the post-disruptive phase of a typical gas injection aided shot 64591. No quantitative statement is made on the runaway current from this ex-vessel hard X-ray detector signal, as it also depends on the RE energy. Note that the sudden increase in PMTX signal at 0.65$\,$s corresponds to the point in time when the vertical plasma position (b) stabilizes, and comes with a 10$\,$kA jump in the plasma current (c).

\begin{figure*}
\centering
\includegraphics[width=0.9\textwidth]{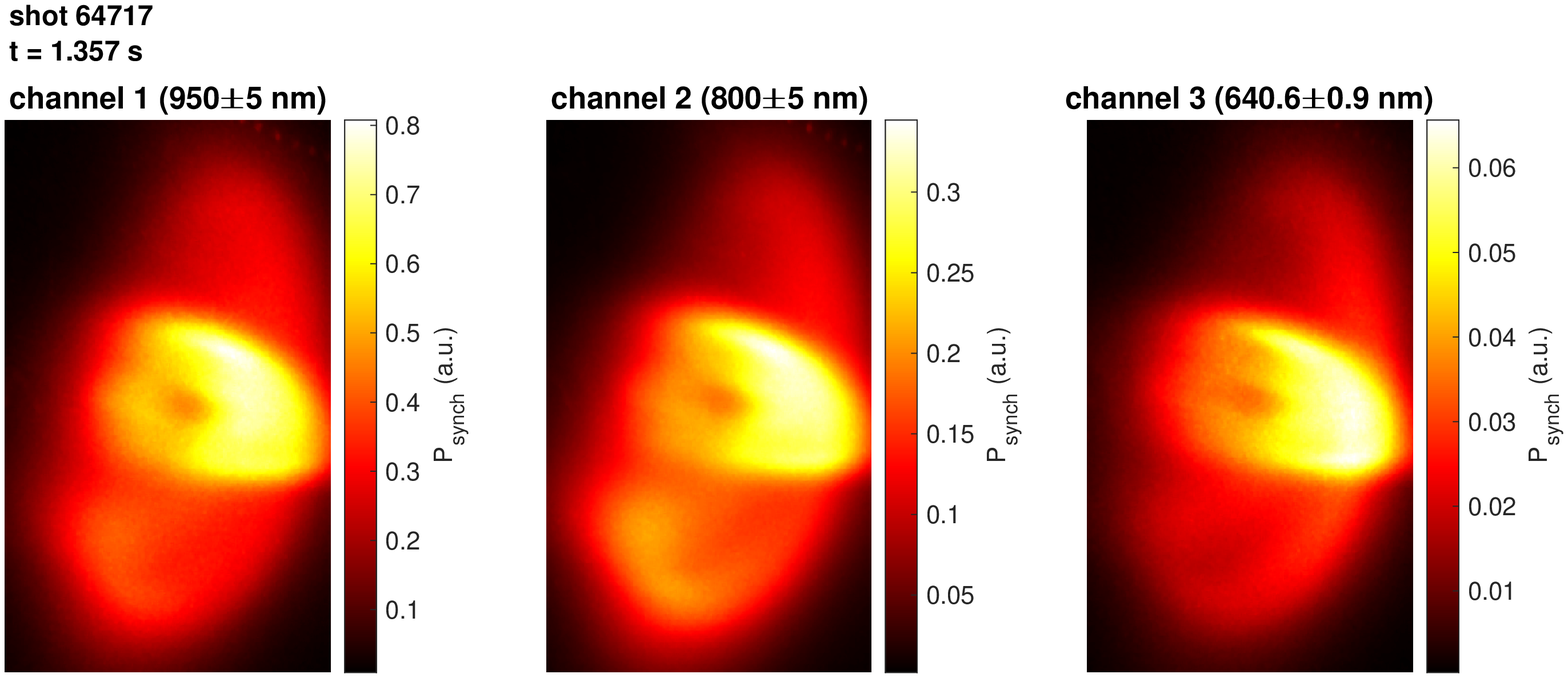}
\caption{Synchrotron radiation patterns observed in TCV at 1.357$\,$s into shot 64717. The filters of channels 1,2 and 3 have central wavelengths at 950, 800 and 640.6$\,$nm respectively.}
\label{fig:4}
\end{figure*}

In a subset of shots characterized by a relatively high loop voltage, light appears on the channels 1-3 as shown in figure \ref{fig:4}. Since (i) the source appears to be continuous, (ii) spectrometers not viewing the runaways head-on do not record the emission and (iii) no line radiation is expected to show up on the first three channels, it can be concluded that it is in fact synchrotron radiation which is observed by MultiCam. This conclusion is reinforced by SOFT simulations, which show patterns highly representative of the experimental results \cite{Hoppe2020}. The integrated synchrotron signals shown in figure \ref{fig:3}f rise until 1.2$\,$s after which the intensity stabilizes, indicating a balance in the mechanism populating the visible region of phase-space with runaway energy and particle losses.

TCV shot 64717 is of a quiescent nature, the low density flat top not being perturbed throughout the shot. Such a state, with limited runaway current fraction, lends itself for characterization of the background plasma with the available diagnostics. For the purpose of assessing the plausibility of the obtained runwaway distribution in the light of the plasma parameter evolution, the quiescent scenario is therefore the preferred one. All synchrotron images shown in this work were taken at 1.357$\,$s in the stable pattern phase. Median filtering is applied to remove X-ray white noise.


\section{Synthetic synchrotron diagnostic SOFT}
\label{sec:ch4}

Before demonstrating a novel reconstruction approach to extract energy, pitch angle and spatial information on the runaways from synchtrotron images, the synthetic synchrotron diagnostic SOFT \cite{Hoppe2018_2} is briefly introduced. This tool bridges the gap between experimentally observed patterns and the runaway distribution function. The code constructs a Green function $G$, also called response function, from a magnetic field geometry and detector details using guiding-center theory. This response function links an arbitrary distribution function $f_{\mathrm{e}}$ in flux surface coordinate $\rho$, momentum $p$ and pitch angle $\theta_{\mathrm{p}}$ to a power distribution $P_{\mathrm{synch}}$ over the specified detector. In this work momentum is normalized to the product of electron mass $m_{\mathrm{e}}$ and the speed of light $c$. In calculating $G(\rho,p,\theta_{\mathrm{p}})$ SOFT takes into account the $(p,\theta_{\mathrm{p}})$ dependence of the emission spectrum. Also it incorporates the highly non-isotropic nature of synchrotron emission, which is emitted along the velocity vector of the runaway. Therefore only light from REs moving parallel to the detector line-of-sight can be observed. The theory underlying the code rests on the assumptions of toroidal symmetry of $f_{\mathrm{e}}$ and the poloidal runaway transit time being much smaller than the runaway collision time. The relation between the power load on pixel $i$ and the RE distribution is:

\begin{equation}
    \label{eq:2}
    \eqalign{
    P_{\mathrm{synch},i} = & \int\int\int\int G_{i}(\rho,p,\theta_{\mathrm{p}},\lambda) f_{\mathrm{e}}(\rho,p,\theta_{\mathrm{p}})
    \\
    &
    {p}^2 \sin(\theta_{\mathrm{p}}) T(\lambda) \mathrm{d{\rho}d{p}d{\theta_{\mathrm{p}}}d{\lambda}}.
    }
\end{equation}

\noindent Here the response function has been weighed by the transmission curve $T(\lambda)$ of the camera filter, and the phase-space Jacobian $J = p^2 \sin(\theta_{\mathrm{p}})$. The magnetic field geometry is obtained from TCV's Grad-Shafranov solver LIUQE \cite{Moret2015}. Note that the availability of a magnetic reconstruction is a necessary condition for composing a response matrix, and is guaranteed by the quiescent nature of the experiments discussed here. The camera location, viewing direction and distortion follow from the Calcam calibration.

Note that at high energies the runaways can experience a significant drift orbit shift, shifting the synchrotron pattern towards the outboard side of the tokamak \cite{Guan2010}. This behaviour is accounted for in SOFT by including a first order correction term to the guiding-center momentum vector as described in recent work by Hoppe et al \cite{Hoppe2020_2}.

Synchrotron radiation patterns are often dominated by the contribution from a narrow part of runaway phase-space. It is the location where the product $GJf_{\mathrm{e}}$ peaks that contributes most to what a camera observes, and this quantity is therefore referred to in this work as the contribution function. After reconstructing a distribution function $f_{\mathrm{e}}$ from an experimental pattern, $GJf_{\mathrm{e}}$ plays an important role in the interpretation of the found $f_{\mathrm{e}}$. Changes in the distribution only affect the image fitting performance there where the contribution function is significant, so that only in this part of phase-space the reconstruction is expected to give an accurate indication of the real distribution. What contribution qualifies as significantly high will be addressed in the next section.


\section{Tomographic reconstruction of the RE distribution from SOFT response functions}
\label{sec:ch5}

Backwards modelling of the runaway population from synchrotron images has to this date relied on analytical or numerical multi-parameter models for the momentum-space distribution \cite{Hoppe2020_2}. This approach effectively tackles the mathematical difficulty of having to estimate the population of all points in 3D phase-space, which amount to a great number of free parameters. Using the constraints imposed by all pixel values in the multispectral images, an attempt is made in this work to let go of the analytical solution and perform a free reconstruction of the runaway distribution. The motivation for retaining this high level of flexibility are unsuccessful first attempts at explaining the TCV synchrotron patterns using kinetic theory \cite{Hoppe2020}. It is therefore not along the line of expectations that the actual distribution will follow a known analytical solution.

\subsection{Tomographic reconstruction method}
\label{sec:ch5.1}

This section will outline the tomographic approach which is adopted for estimating the solution to the system of equations linking the distribution to the image via the response function:

\begin{equation}
    P_{\mathrm{synch},i} = \sum_{j}(GJ)_{ij} f_{\mathrm{e},j},
\end{equation}

\noindent where $(GJ)_{ij}$ is the transmission corrected product of the response function and Jacobian for pixel $i$ and phase-space index $j$, the latter referring to a specific point in $(\rho,p,\theta_{\mathrm{p}})$ space. The equations for the pixels in all available image channels together form a linear system of equations, for which the distribution solution is approximated using the simultaneous algebraic reconstruction technique (SART) \cite{Andersen1984,Carr2018}. SART is an iterative way of converging towards a solution minimizing the difference between the synthetic image $GJf_{\mathrm{e}}$ and the real image $P_{\mathrm{synch}}$, using the difference between the two to update all points in the distribution simultaneously:

\begin{equation}
\label{eq: SART}
    f_{\mathrm{e},j}^{l+1} = f_{\mathrm{e},j}^{l} + \Delta f_{\mathrm{e},j|\mathrm{SART}}^l + \Delta f_{\mathrm{e},j|\beta_{\mathrm{m}}}^l + \Delta f_{\mathrm{e},j|\beta_{\mathrm{\rho}}}^l.
\end{equation}

\noindent The distribution increment in each step $l$ first of all depends on the SART contribution itself:
    
\begin{equation}
    \eqalign{
    \Delta f_{\mathrm{e},j|\mathrm{SART}}^l = & \frac{\lambda_{\mathrm{r}}}{\sum_{i}(GJ)_{ij}} \cdot
    \\
    &
    \sum_{i} \frac{(GJ)_{ij}}{\sum_{j}(GJ)_{ij}} 
    \Big( P_{\mathrm{synch},i} - (GJ)_{ij} f_{\mathrm{e},j}^{l} \Big)
    }
\end{equation}

\begin{figure*}
\centering
\includegraphics[width=1\textwidth]{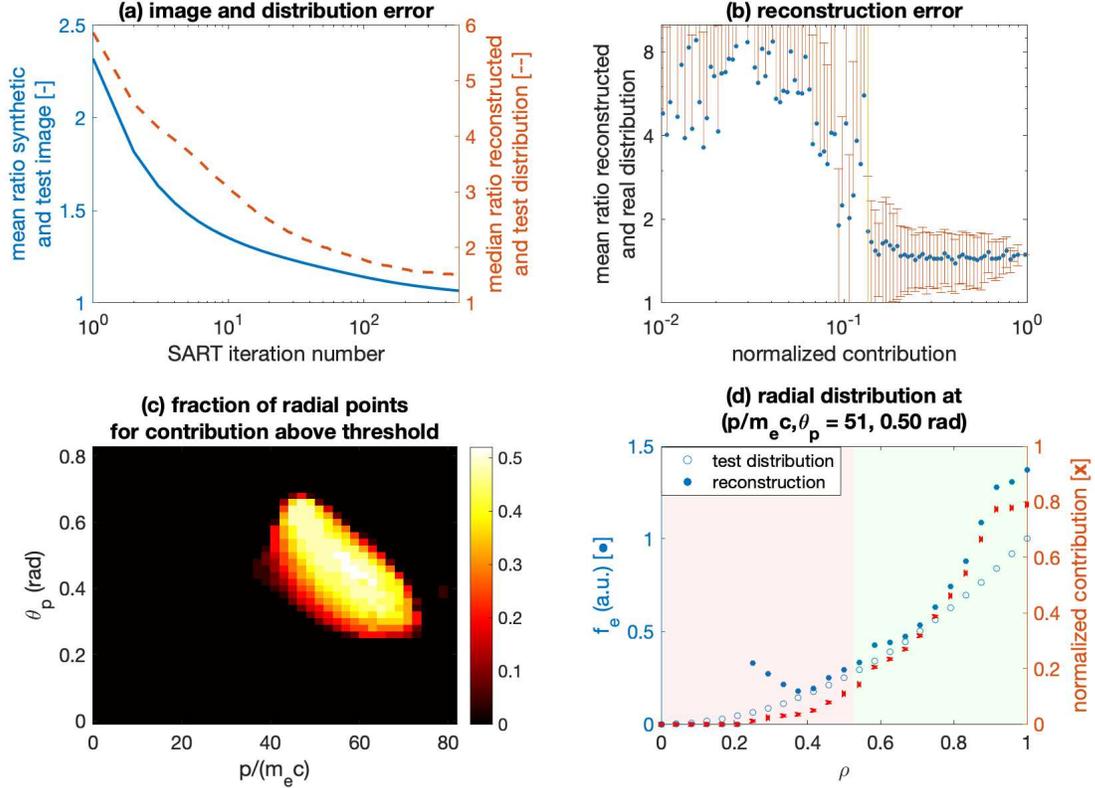}
\caption{Accuracy assessment of the SART reconstruction approach for an avalanche-like distribution function. No smoothing is applied in the 500 SART iterations used for the reconstruction. a) Evolution of the mean image and median distribution error. For the image comparison, only pixels with an intensity larger than 1 percent of the maximum intensity are examined. Distribution points with a contribution function $GJf_{\mathrm{e}}$ value larger than 10 percent of the maximum are considered. b) Ratio between the reconstructed and input distribution as a function of contribution of the reconstructed distribution to the total image. The definition of the ratio is such that the largest of the two is divided by the smallest, so that 1 indicates perfect resemblance. c) Momentum-space map of the fraction of flux surface points above the normalized contribution threshold of 0.135. The threshold is based on the data in figure b. d) For a normalized momentum $p/{m_{\mathrm{e}}}c$ of 51, and a pitch angle of 0.50$\,$rad, the test distribution and its reconstruction are depicted as a function of normalized minor radius. The right axis shows the normalized contribution function $GJf_e$. The red shaded region indicates where the contribution drops below 0.135, and the reconstruction is expected to be less accurate than a factor of 2. Note that below $\rho \sim 0.4$ the contribution is negligible, and large deviations between the reconstructed and real distribution start to become apparent.}
\label{fig:5}
\end{figure*}
    
\noindent The pre-factor to the sum is the relaxation term which governs the speed of convergence, and is modified through the relaxation parameter $\lambda_{\mathrm{r}}$. Choosing $\lambda_{\mathrm{r}}$ too large will cause the solution to overshoot, and the starting point recommended here is just below 1. The SART modification is based on the difference between all synthetic and experimental pixel values, the term between brackets, weighed by the response function for the given pixel and distribution element. The distribution increments $\Delta f_{\mathrm{e},j|\beta_{\mathrm{m}}}^l$ and $\Delta f_{\mathrm{e},j|\beta_{\mathrm{\rho}}}^l$ impose a certain smoothness on the solution in momentum- and real-space respectively. This is achieved by invoking a penalty based on the population difference between a point in phase-space and its neighbours:
    
\begin{equation}
    \Delta f_{\mathrm{e},j|\beta_{\mathrm{m}}}^l = - \beta_{\mathrm{m}} \Big( N_{n,\mathrm{m}} f_{\mathrm{e},j}^{l} - \sum_{n=1}^{N_{n,\mathrm{m}}} f_{\mathrm{e},n}^{l} \Big),
\end{equation}

\begin{equation}
    \Delta f_{\mathrm{e},j|\beta_{\mathrm{\rho}}}^l = - \beta_{\mathrm{\rho}} \Big( N_{n,\mathrm{\rho}} f_{\mathrm{e},j}^{l} - \sum_{n=1}^{N_{n,\mathrm{\rho}}} f_{\mathrm{e},n}^{l} \Big).
\end{equation}

\noindent Here $N_{n,\mathrm{m}}$ is the number of neighbours on the $(p,\theta_{\mathrm{p}})$ grid, and $N_{n,\mathrm{\rho}}$ the number of neighbours in the radial direction. The index $n$ refers to the distribution index of these adjacent elements of $f_{\mathrm{e}}$. Note that the imposed smoothness is controlled separately for momentum- and real-space through the parameters $\beta_{\mathrm{m}}$ and $\beta_{\mathrm{\rho}}$.

The SART algorithm is terminated after a predefined number of steps, or when the relative change in the total distribution;

\begin{equation}
\label{eq: SART termination}
    \sum_{p,\theta_{\mathrm{p}}}|f_{\mathrm{e}[p,\theta_{\mathrm{p}}]}^{l}-f_{\mathrm{e}[p,\theta_{\mathrm{p}}]}^{l-1}|/\sum_{p,\theta_{\mathrm{p}}} f_{\mathrm{e}[p,\theta_{\mathrm{p}}]}^{l-1},
\end{equation}

\noindent reaches a specified threshold. The influence of the smoothness operators on the solution accuracy and convergence speed is assessed later in this section.

\subsection{Accuracy assessment of the SART method}
\label{sec:ch5.2}

To assess how the accuracy and convergence speed of the SART algorithm depend on the solver parameters $\lambda_{\mathrm{r}}$, $\beta_{\mathrm{m}}$ and $\beta_{\mathrm{\rho}}$, a test distribution is constructed. An analytical avalanche distribution of the form \cite{Fulop2006}:

\begin{equation}
    \label{eq: avalanche equation}
    \eqalign{
    f_{\mathrm{e}} = & \frac{C_0 (\rho)}{\cos(\theta_{\mathrm{p}}) p} \exp \left\{ C_1(\rho) \cos(\theta_{\mathrm{p}}) p \right\}
    \\
    &
    \exp \left\{ C_2(\rho) \tan(\theta_{\mathrm{p}}) \sin(\theta_{\mathrm{p}}) p \right\},
    }
\end{equation}

\noindent is used, where the coefficients $C_{0,1,2}$ all depend on the flux surface coordinate. From the SOFT response functions for TCV shot 64717 at 1.357$\,$s, the camera images for channels 1 through 3 are generated. These images are then perturbed using a pixel intensity dependent noise level, and fed to the SART algorithm. This allows one to compare the reconstructed distribution with the actual one for distributions of various complexities. It is important to note that this accuracy assessment does not cover reconstruction errors introduced by assumptions underlying the guiding-center framework used in SOFT.

For all simulations performed in this work the response functions are generated on a $41 \times 42$ momentum, pitch angle grid ranging from $(p/{m_{\mathrm{e}}}c,\theta_{\mathrm{p}}) = (1, 0 \,$rad) to $(81, 0.82 \,$rad). The original flux surface resolution is 100 to ensure smooth synthetic patterns, but binned into 25 points to reduce memory demands. For the same reason experimental images are binned 8 by 8 to reduce the response function size.

\subsubsection{Smooth radially dependent profile}
\label{sec:5.2.2}
\hfill

To approach an experimentally realistic distribution, a flux surface dependent momentum-space profile is used. This is achieved through a $\rho$ dependent value of the avalanche coefficients. $C_{0,1,2}$ are tuned so as to generate a synchrotron patterns resembling those in figure \ref{fig:4}. $C_2$ is left at a constant $-0.3$, while $C_1$ decreases linearly from $-0.08$ to $-0.14$ between the center and edge of the plasma. $C_0$ is chosen in a way that assures that $f_{\mathrm{e}} = \rho^2$ at a normalized momentum of 51, and pitch angle of 0.50$\,$rad. A discussion of this unconventional radial profile is deferred to section \ref{sec:ch6.4}. Intensity dependent noise is added by scaling all intensities between 0 and 4096, matching the camera sensor signal levels, and then taking the square root of each pixel as its standard deviation. From each pixel value distribution an arbitrary integer is drawn.

For a relaxation parameter of $\lambda_{\mathrm{r}} = 0.95$ and no smoothing, the evolution of the mean ratio between the pixel values of the test- and synthetic image as a function of SART iterations is depicted in figure \ref{fig:5}a. In computing the mean, only pixels with an intensity above a fraction of 0.01 of the test image maximum have been considered. The pixel ratio $R_i$ has been defined such that the largest value, from the synthetic or test distribution, is divided by the smallest one:

\begin{equation}
    \label{eq: ratio error}
    R_i = \mathrm{max} \left[ \frac{P_{\mathrm{synch},i-\mathrm{synthetic}}}{P_{\mathrm{synch},i-\mathrm{test}}}, \frac{P_{\mathrm{synch},i-\mathrm{test}}}{P_{\mathrm{synch},i-\mathrm{synthetic}}} \right].
\end{equation}

\noindent A perfectly reconstructed pixel value therefore corresponds to a value of 1. This error measure weighs an underestimation by a factor of 2 the same as an overestimation by a factor of 2. Similarly, a ratio based error measure for the distribution function can be defined. Figure \ref{fig:5}a depicts the median of this ratio for phase-space elements which have a contribution value $GJf_{\mathrm{e}}$ above 10 percent of the peak value. This contribution function is based on the test distribution.

The image error is a strictly decreasing function of the number of iterations. After 500 iterations, taking roughly 5 minutes on an Intel i9 core, the relative error is reduced to 6.6 percent. Throughout the convergence process, all pixel errors decrease in a similar fashion. The convergence is not limited to a specific part of the image. The main quantity of interest is however the distribution function error. As shown, the distribution error follows the decreasing trend. After the indicated number of iterations the median of the relative errors is 50 percent. This error should be put in the context of the $f_{\mathrm{e}}$ spread in the considered region of phase-space which spans 4 orders of magnitude. The steep decrease in $f_{\mathrm{e}}$ with momentum and pitch angle is captured well by the reconstruction method.

An important question to ask is what part of the reconstructed distribution is to be trusted. In figure \ref{fig:5}b the distribution ratio error and its standard deviation are plotted against the normalized contribution of each phase-space element to the image. Normalization is with respect to the element of $f_{\mathrm{e}}$ contributing most to the perceived intensity. A clear drop in the error is observed above a $GJf_{\mathrm{e}}$ value of 0.135. The location of the corresponding points is visualized in figure \ref{fig:5}c. Reconstructions are clearly only to be trusted there where the distribution has a significant contribution to the image. The synthetic image is not sensitive to features elsewhere in the distribution and the synchrotron footage can therefore not be used to constrain all of momentum- and real-space. 

The contribution is not just a function of momentum, but also of flux surface coordinate. It is the edge of the distribution which dominates the synthetic image, and large deviations between the real and reconstructed distribution are found near the core of the plasma. To illustrate this, the distribution is plotted as a function of $\rho$ for the momentum-space point $(p/{m_{\mathrm{e}}}c,\theta_{\mathrm{p}}) = (51, 0.50 \,$rad) in figure \ref{fig:5}d. The reconstruction ratio error is below 37 percent for $\rho>0.4$. Further inwards, the contribution function drops to zero and large deviations from the test distribution are observed.

What level of $GJf_{\mathrm{e}}$ is sufficient for a reconstruction from experimental images to be trusted remains undetermined. The accuracy assessment does indicate that the SART algorithm can locally obtain results accurate to within a factor of 2, spanning over a considerable fraction of phase-space. Equally important, the analysis points towards a prominent role of the contribution function in interpretation of a reconstructed distribution. Lastly, figure \ref{fig:5} reveals a clear trade-off between accuracy and speed. Time dependent reconstructions covering several tens of frames are however well within reach, despite little time having been devoted to speed optimization of the analysis framework.

\subsubsection{Flux surface dependent energy cut-off}
\label{sec: 5.2 disc}
\hfill

\begin{figure}
\centering
\includegraphics[width=0.40\textwidth]{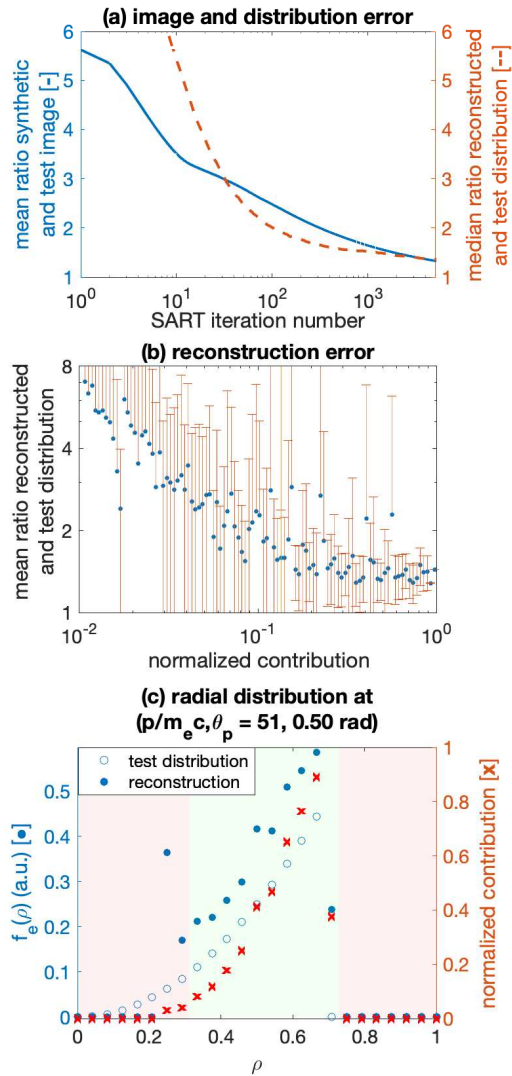}
\caption{Accuracy assessment for the SART method. The distribution is of an avalanche form, but all runaways with normalized momentum larger than 30 at $\rho>0.7$ have been removed.  a) Evolution of the thresholded mean image and median distribution error. b) Ratio between the reconstructed and input distribution as a function of contribution of the reconstructed distribution to the total image. c) Radial distribution at a fixed momentum and pitch angle, showing the reconstructed radial discontinuity after 6000 SART iterations. The part of the distribution for which the reconstruction error is expected to be lower than a factor of 2 is indicated in green.}
\label{fig:6}
\end{figure}

\begin{figure*}
\centering
\includegraphics[width=1.0\textwidth]{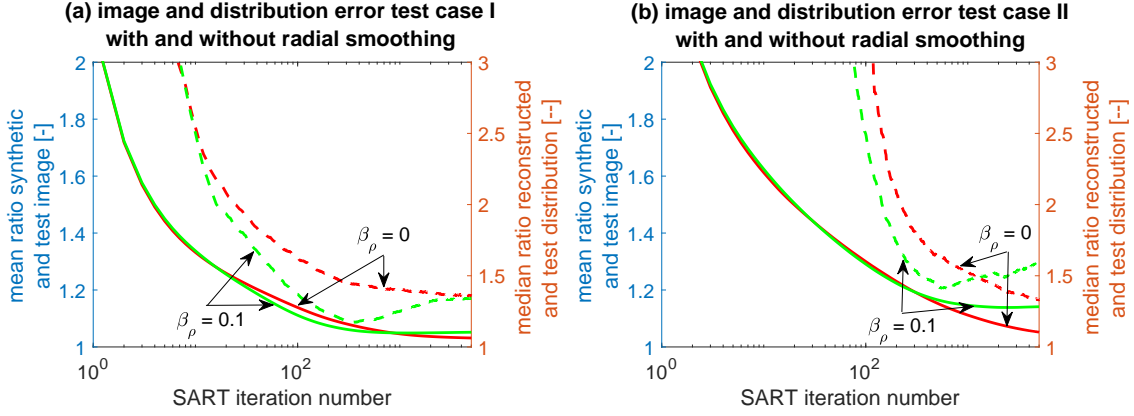}
\caption{Image and runaway distribution reconstruction as a function of the number of SART iterations for no smoothing ($\beta_{\mathrm{\rho}}$ = 0) and finite smoothing at a level of $\beta_{\mathrm{\rho}}$ = 0.1. a) Performance for the distribution discussed in section \ref{sec:5.2.2}. b) Performance for the distribution discussed in section \ref{sec: 5.2 disc}.}
\label{fig:7}
\end{figure*}

The limits of the reconstruction method will be tested using a jump in the distribution. For this test case it is assumed that at the edge of the plasma, for $\rho>0.7$, efficient energy blocking occurs at a normalized momentum of 30. The result is a sudden jump in both momentum-space and radial direction. Such energy blocking could occur due to the presence of externally injected high frequency waves at the edge of the plasma \cite{Guo2018}. Apart from the absence of runaways at $(p/({m_e}c),\rho)>(30,0.7)$, the distribution remains the same as for the previous test case.

Performing the SART reconstruction, it is found that the image and distribution convergence are slower than for the distribution studied in the previous section. As demonstrated in figure \ref{fig:6}a, the median distribution ratio between the reconstructed and test distribution continues to decrease well beyond 500 iterations. This indicates that convergence time will in general depend on the case at hand, and calls for a SART termination criterion based on the relative change of the distribution.

The ratio error as a function of the contribution function after 6000 iterations is depicted in figure \ref{fig:6}b, and shows the same behaviour as for the more continuous distribution. The SART algorithm appears capable of handling the steep spatial barrier. This statement is backed up by the radial profile at $(p/{m_{\mathrm{e}}}c,\theta_{\mathrm{p}}) = (51, 0.50 \,$rad) shown in figure \ref{fig:6}c. At $\rho = 0.7$ a sudden drop in the reconstructed population is observed, matching the location of the barrier in the real distribution. There is a transition region between the occupied and empty part of real-space of a single radial point. It is important to note that the distribution features at $\rho > 0.7$ are the ones converging after 500 iterations. In order to resolve distributions with steep gradients it is therefore of paramount importance to study the behaviour of the distribution with the number of SART iterations before interpreting the profile.

In conclusion, it has been shown that the unconstrained SART method is able to retrieve the test distribution with a relative accuracy of a few tens of percents for all elements contributing significantly to the synthetic image. To stay below a relative error of a factor of 2, typically only elements with a contribution at a level above 10 percent of the maximum contribution should be considered for the discussed experimental setup.

\subsubsection{Influence of smoothing parameters}
\hfill

The influence of tuning the smoothness parameters $\beta_{\mathrm{\rho}}$ and $\beta_{\mathrm{m}}$ is evaluated using the distribution functions studied in sections \ref{sec:5.2.2} and \ref{sec: 5.2 disc}. First the radial smoothing is addressed. A 5000 iteration SART reconstruction is performed for various values of $\beta_{\mathrm{\rho}}$. Common to all cases with finite smoothing is that the error in the reconstructed image and distribution reach a minimum after some number of iterations. For a $\beta_{\mathrm{\rho}}$ value of 0.1, this is shown in figure \ref{fig:7}. The number of iterations after which the convergence starts to degrade decreases with increasing smoothing. Smoothing beyond a value of 0.5 yields a non-converging solution. For the test case presented in section \ref{sec:5.2.2}, smoothing at a level of 0.1 can improve the achieved distribution reconstruction accuracy as shown in figure \ref{fig:7}a. The difficulty is however in finding the right number of SART iterations when performing reconstructions from experimental data. The minimum in image and distribution error do not coincide, so the former cannot be used to formulate a termination criterion.

For the case of the distribution with large gradients, as introduced in section \ref{sec: 5.2 disc}, the performance is displayed in figure \ref{fig:7}b. The SART algorithm is not able to reconstruct the spatial barrier when smoothing is applied. The jump in distribution at $\rho = 0.7$, shown in figure \ref{fig:6}b, is not recovered due to the penalty invoked on the presence of this feature.

Applying smoothing in momentum-space degrades the algorithm performance regardless of the number of iterations and the level of smoothing. This is found for both test distributions, and leads to the strong recommendation not to use smoothing in the $p$ and $\theta$ dimensions.

In conclusion, smoothing should only be applied in the radial direction and stay at a level of 0.5 or lower. Before using smoothing, one should first confirm that the non-smoothed reconstruction does not show signs of large gradients. Lastly, the number of SART iterations should stay below the number at which a minimum in the image error is found.

\subsection{Added value of multispectral imaging}

\begin{figure*}
\centering
\includegraphics[width=0.9\textwidth]{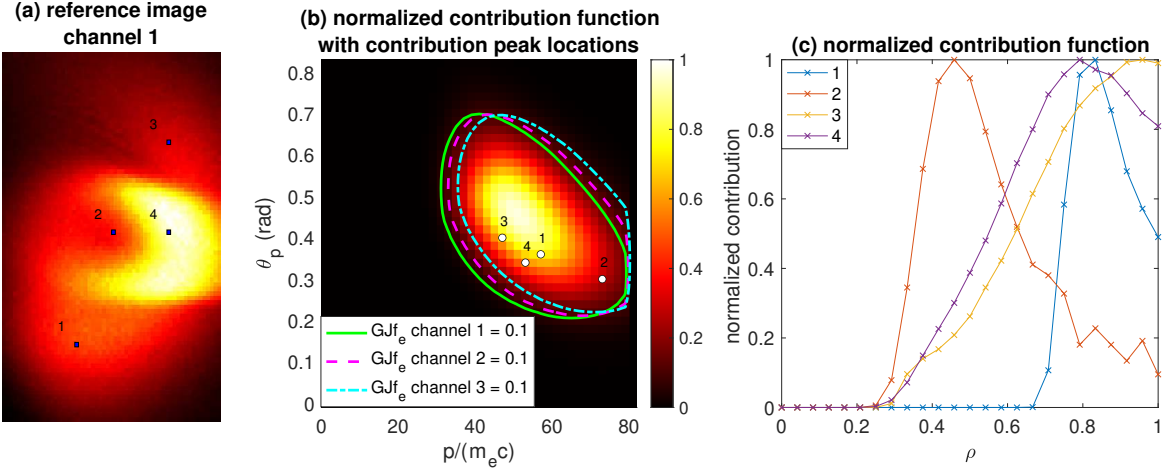}
\caption{Spectral and spatial dependence of the contribution function for the avalanche-like distribution from section \ref{sec:ch5.2}. a) Binned synchrotron image associated with the distribution. 4 points are labelled, for which the peak in the contribution function is indicated in figure b. b) Normalized contribution function averaged over the three channels. The colored contours indicate the $GJf_{\mathrm{e}} = 0.1$ boundary, and show that the probed region of momentum-space moves towards higher momenta for smaller central filter wavelengths. c) For the 4 points in the image the average flux surface contribution is shown. Different parts of the image contain information on different parts of the spatial distribution.}
\label{fig:8}
\end{figure*}

As shown, distribution reconstructions are only accurate where the phase-space contribution to the image is sufficiently large. The merit of the distribution reconstruction method therefore depends on the extent of the phase-space region to which the synchrotron diagnostic is sensitive. In this section it is demonstrated that the spectral- and spatial resolution of a multispectral camera systems achieve a complementary widening of the probed momentum- and real-space.

To illustrate the point, the test distribution from section \ref{sec:5.2.2} is used. The corresponding synthetic synchrotron image is shown in figure \ref{fig:8}a. In the image four different pixel clusters are highlighted. Each cluster has its own associated contribution function $GJf_{\mathrm{e}}$, the peak values of which are indicated in figure \ref{fig:8}b. Figure \ref{fig:8}b also displays the normalized contribution function for the image as a whole, averaged over the three camera channels. Note from the four peak locations that different parts of the image predominantly constrain different parts of the runaway momentum-space. 

The flux surface dependence of the contribution functions is portrayed in figure \ref{fig:8}c. It shows that positions 1, 3 and 4 mainly receive light from runaways at the edge of the plasma, while position 2 is sensitive to the distribution closer to the core. The peak positions in momentum-space are tied to these observations on the radial contribution profiles. Moving from the center of the plasma to the edge, the RE pitch angle required for the camera line-of-sight to be parallel to the runaway velocity vector increases. Therefore, the pixels which probe the edge of the distribution are expected to be more sensitive to higher pitch angles, as is indeed observed in figure \ref{fig:8}b.

Next, consider the effect of the central filter wavelength on the probed region of phase-space. The normalized $GJf_{\mathrm{e}} = 0.1$ bounds for the three different channels are indicated in figure \ref{fig:8}b. It is found that channels with a shorter central wavelength probe a higher $(p,\theta_{\mathrm{p}})$ region of momentum-space. This is a direct consequence of the fact that the synchrotron emission peak moves from the infrared towards the visible with increasing momentum and pitch angle. Thus, the contribution from the REs with highest $(p,\theta_{\mathrm{p}})$ to the total image increases with decreasing wavelength. A larger spread in the probed momentum-space can be achieved by increasing the separation in filter central wavelength. For MultiCam and other visible light cameras the significant line emission of hydrogen and regularly encountered impurities below $\sim$ 600$\,$nm limit the possible extend of such an optimization.

Interestingly, extending the range of considered wavelengths broadens the ellipse-like contribution peak region along its minor axis. The sensitivity spread accomplished by adding the spectral dimension to 2D imaging is thus complementary to that achieved by the spatial resolution itself. Multispectral imaging systems essentially act as a large collection of absolutely calibrated spectrometers with coarse wavelength resolution and known spatial orientation, providing sufficient coverage of phase-space to make the SART reconstruction possible.

To quantify the increase in reconstruction performance by having multiple image channels, the image and distribution accuracy for SART input from a single channel are compared to those for input from all three channels. Allowing both reconstructions to last 100$\,$s, the single channels approach yields a smaller mean image ratio error of 1.06, as compared to the 1.11 for the three channels approach. Apparently, with the increased number of equations to satisfy, finding an exact match to the image data is complicated by using multiple channels. The distribution error however shows the reverse trend. Using a single channel yields a mean ratio error of 2.13, while this is only 1.64 for the three channel case. As distribution reconstruction is the sole aim of the SART algorithm, this increase in performance is the leading argument to favour multiple channel reconstruction over the use of a single image.

\section{Origin of TCV synchrotron pattern}
\label{sec:ch6}

\subsection{Experimental reconstruction results}

Using the SART algorithm an attempt is made to estimate the runaway distribution corresponding to the TCV synchrotron images displayed in figure \ref{fig:4}. Changes to the image error and the distribution function are marginal after 20 iterations. This shorter convergence time as compared to the test case is attributed to simplifications made in the model underlying the synthetic diagnostic. As a result, one cannot expect perfect convergence of the synthetic image to the experimental one, and the best approximation to the image is reached in fewer iterations. To allow for fast time sequence reconstructions, the number of iterations is therefore not increased beyond 20. The relaxation parameter $\lambda_{\mathrm{r}}$ is set at 0.95 which, under the constraint of getting a converging solution, maximizes the algorithm speed. It is found that setting the radial smoothing parameter $\beta_{\mathrm{\rho}}$ to 0.1 the image error reaches a minimum after 20 iterations, after which the error starts increasing rapidly. To make sure smoothing does not inadvertently introduce artefacts in the reconstruction, it is turned off by setting $\beta_{\mathrm{\rho}}$ to 0. For the SART input, all three camera channels are used. Also for the reconstruction of $f_{\mathrm{e}}$ from experimental images it is found that the image error is twice as large compared to the single image reconstruction. Based on the results from the test distribution in the previous section, using all channels is however expected to yield a more accurate distribution result, which is the main aim of the analysis presented here.

\begin{figure}
\centering
\includegraphics[width=0.48\textwidth]{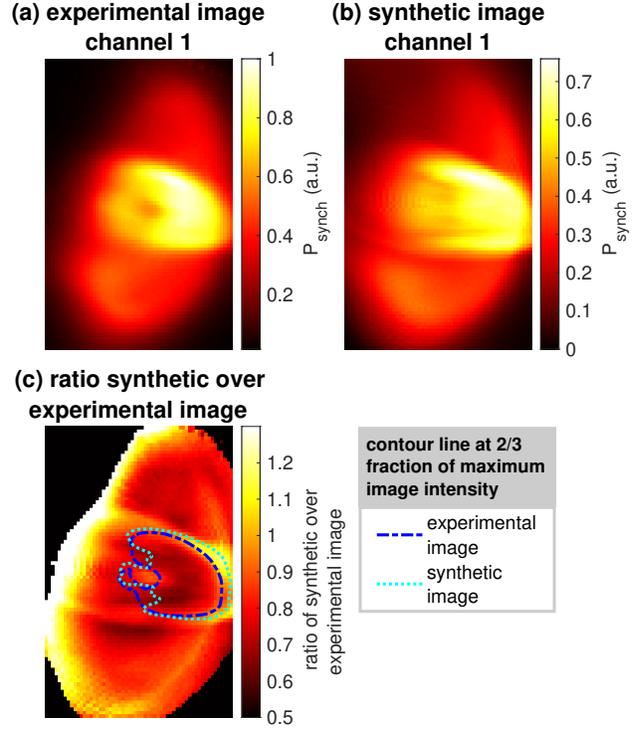}
\caption{Synchrotron image reconstruction results for TCV shot 64717 at 1.357$\,$s. a) Experimental image for channel 1. b) Corresponding synthetic image after 20 SART iterations. c) Ratio between the synthetic and experimental image for pixels with an intensity above 1 percent of the maximum. The contour lines indicate the position where the image is at a $2/3$ fraction of its maximum, and reveal a spatial offset of the synthetic image with respect to the experimental one.}
\label{fig:9}
\end{figure}

\begin{figure*}
\centering
\includegraphics[width=1.0\textwidth]{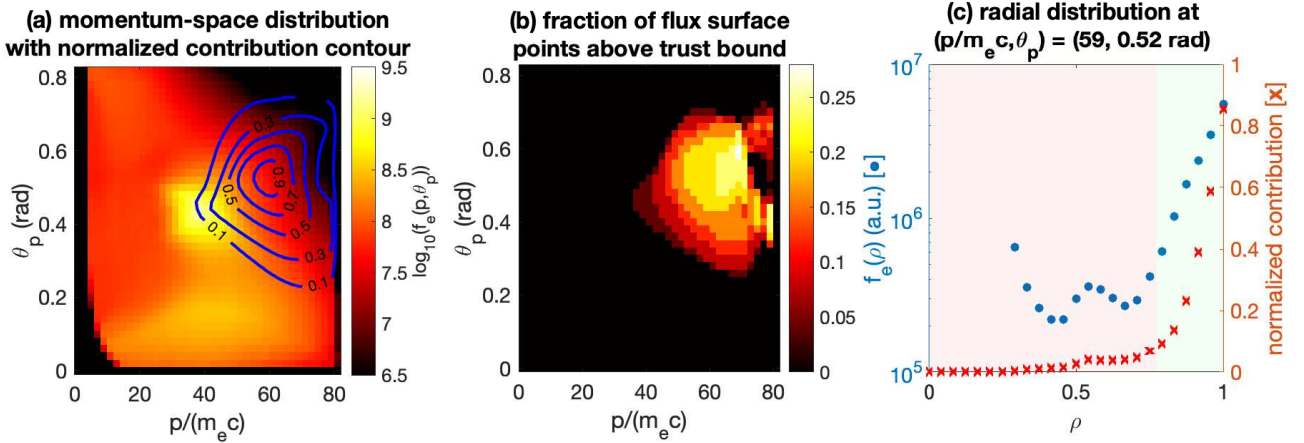}
\caption{Runaway distribution reconstruction results for TCV shot 64717 at 1.357$\,$s. a) Logarithm of runaway distribution in momentum-space, integrated over all flux surfaces. The contour lines indicate the normalized value of the contribution function $GJf_{\mathrm{e}}$. b) Number of flux surface points above a normalized contribution threshold of 0.1. c) Logarithm of flux surface distribution at $(p/{m_{\mathrm{e}}}c,\theta_{\mathrm{p}}) = (59, 0.52 \,$ rad). The normalized contribution function at each point is indicated. Only results in the green shaded region are expected to approach the real distribution.}
\label{fig:10}
\end{figure*}

The reconstruction results at time 1.357$\,$s into TCV shot 64717 are portrayed in figures \ref{fig:9} and \ref{fig:10}. In figure \ref{fig:9} the experimental and synthetic image are compared. Both the overall pattern shape and the intensity distribution are comparable. The integrated intensity differs by a few tens of percent, the mean image ratio error above the 1 percent intensity threshold dropping to 1.65 after 20 SART iterations. The spatial distribution of the ratio between the two images is displayed in figure c, where only pixels with an intensity above 1 percent of the maximum are studied. Contour lines indicate the location where the pattern for the experimental and synthetic case are at a fraction of $2/3$ of their maximum, and reveal a spatial offset. The error is most pronounced where the lobe overlap regions, yellow in figures a and b, of the synthetic and experimental image do not coincide.

The reconstructed momentum-space distribution integrated over all flux surfaces is shown in figure \ref{fig:10}a. The contour lines indicate the contribution function normalized to its maximum value. It shows the region of maximum contribution is at high pitch angle and momentum, with the peak at $(p/{m_{\mathrm{e}}}c,\theta_{\mathrm{p}}) = (60, 0.53 \,$rad). The momentum-space profile is discussed in more detail in section \ref{sec:ch6.3}

Figure \ref{fig:10}b shows the fraction of flux surface points satisfying the criterion $GJf_{\mathrm{e}}/\mathrm{max}(GJf_{\mathrm{e}}) > 0.07$ at each point in momentum-space. This bound was chosen by comparing the total image contribution below some level of $GJf_{\mathrm{e}}$ to the accumulated image uncertainty. The map indicates the region of momentum-space where reconstructions are expected to yield accurate results to within a few tens of percent. Note that for a given energy and pitch angle, the distribution shape is only to be trusted for a sub selection of radial points.

For one of the points in momentum space contributing strongly to the image, namely $(p/{m_{\mathrm{e}}}c,\theta_{\mathrm{p}}) = (59, 0.52 \,$rad), the radial profile is shown in figure \ref{fig:10}c. At high momenta and pitch angles the distribution is an increasing function of $\rho$.  Mechanisms which could lead to such a profile are discussed in section \ref{sec:ch6.4}.

\subsection{Reconstruction uncertainty}

Besides the systematic errors in the guiding-center theory and reconstruction approach, the estimated distribution has an uncertainty associated with (i) uncertainties in the input image and (ii) uncertainties in the camera calibration. Since the reconstruction method does not specify a direct relation between the input image and output distribution, it is not clear cut how the errors propagate to $f_{\mathrm{e}}$. Here, a Monte Carlo approach is used to quantify the effect of both sources. 

\begin{figure}
\centering
\includegraphics[width=0.40\textwidth]{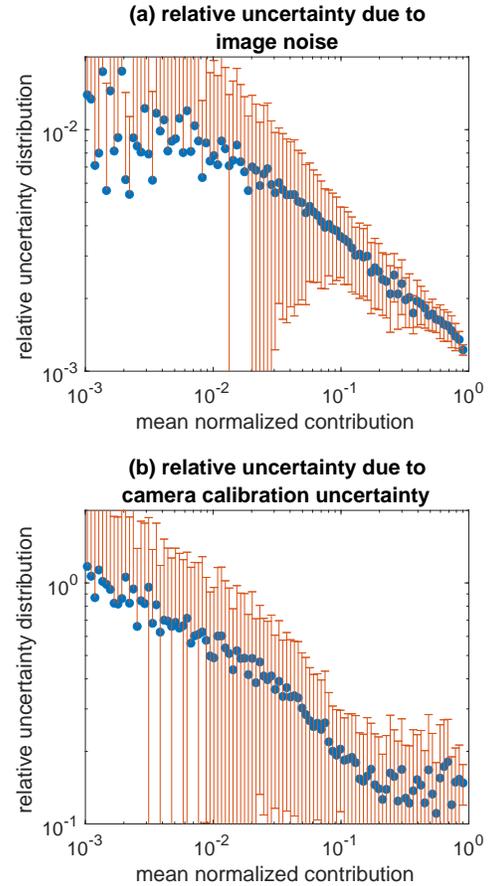}
\caption{Relative uncertainty in distribution reconstruction as a function of contribution level for shot 64717 at 1.357$\,$s. a) Uncertainty due to noise in the input image. b) Uncertainty based on Calcam camera calibration residual pixel error.}
\label{fig:11}
\end{figure}

As mentioned in section \ref{sec:ch2}, experimental image uncertainties are known on a pixel to pixel basis. This allows one to add random noise to the original image by drawing a value from the distribution function of each individual pixel. Repeating this process $N_{\mathrm{image}}$ times for all camera channels a set of images is generated which complies with the experimental uncertainty. By feeding all images to the SART algorithm one-by-one, $N_{\mathrm{image}}$ distribution estimates are obtained. Together, these allow one to construct a mean and standard deviation for the distribution at each point in phase-space. Thus, the propagation of uncertainties in the experimental image to the reconstruction can be quantified. 

Figure \ref{fig:11}a shows the relation between the mean normalized contribution function $GJf_{\mathrm{e}}$ and the standard deviation in $f_{\mathrm{e}}$. Better constrained parts of the distribution suffer less from measurement noise. The overall level of uncertainty for $GJf_{\mathrm{e}}$ above 1 percent of the maximum is of the same order as the pixel value uncertainties. The latter are damped by the pixel binning procedure applied.

Using the same approach the uncertainty introduced by the camera calibration is assessed. Calcam calibrations for MultiCam in TCV typically have a 2 pixel root-mean-square projection error for the calibration points. By applying a horizontal and vertical translation with a value drawn from a distribution with mean 0 and standard deviation of 2 for all three images independently, again a set of $N_{\mathrm{image}}$ images are obtained and used for reconstruction. The results are depicted in figure \ref{fig:11}b. A similar trend as for the image noise is observed, but the relative uncertainties are roughly two orders of magnitude larger. Above the 10 percent of maximum threshold for the contribution function, relative uncertainties are of the order of 10 to 20 percent.

\begin{figure}
\centering
\includegraphics[width=0.41\textwidth]{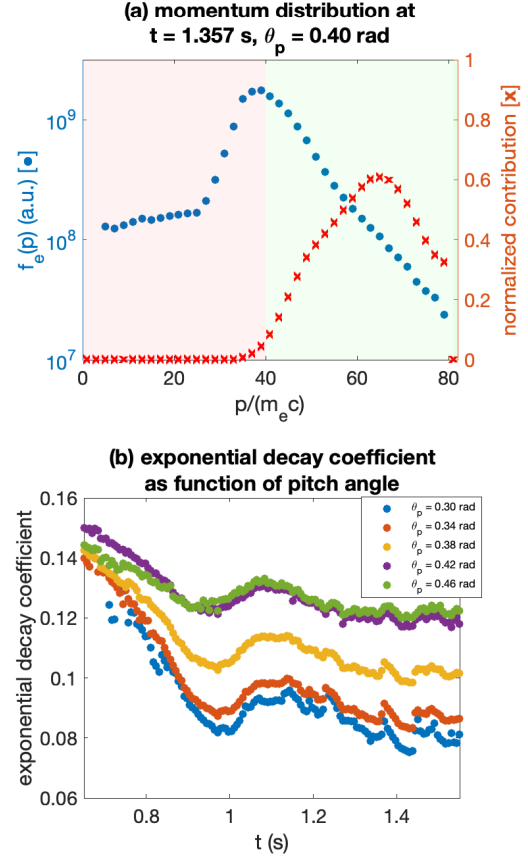}
\caption{Assessment of shot 64717 momentum runaway profile as obtained using the SART algorithm. a) Logarithm of the flux surface integrated distribution $\sum_{\rho} f_{\mathrm{e}}$ as a function of momentum $p$, for a pitch angle of 0.40$\,$rad at time $t = 1.357 \,$s. Decay of $f_{\mathrm{e}}$ appears to be of an exponential nature for sufficiently high contribution values. b) For different pitch angles a linear fit is used to obtain the slope of the exponential decay for the momentum-space points having a contribution value above a fraction of 0.1 of the maximum.}
\label{fig:12}
\end{figure}

\begin{figure}
\centering
\includegraphics[width=0.43\textwidth]{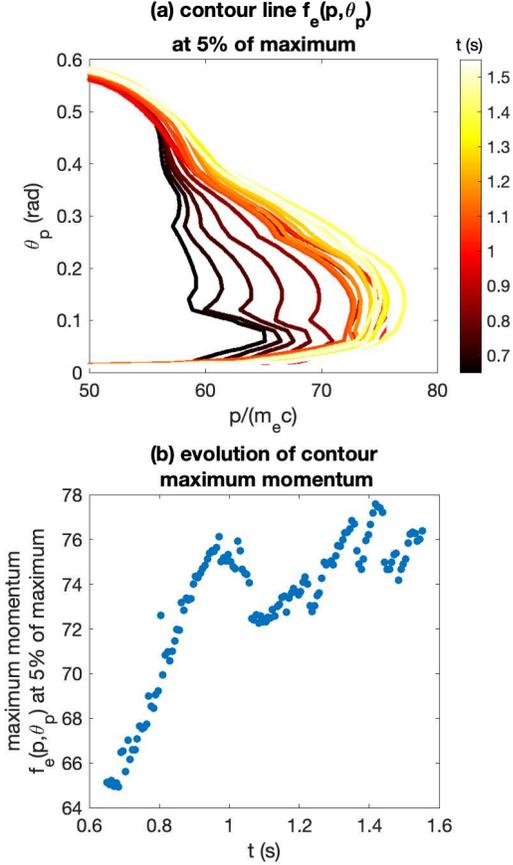}
\caption{Time evolution of the contour line at 5 percent of the maximum value of the flux surface integrated runaway distribution for shot 64717. a) Time dependence of the contour location. The distribution shifts towards higher momenta until 1.2$\,$s. Whereas the front starts out at a nearly uniform value for all pitch angles, the stable momentum location is higher for lower pitch angles. b) Evolution of the maximum momentum value for the contour.}
\label{fig:13}
\end{figure}

It is concluded that errors introduced by reconstruction imperfections, as discussed in chapter \ref{sec:ch5}, are the dominant source of uncertainty. The camera calibration error is however non-negligible. For estimating its magnitude the Monte Carlo uncertainty analysis procedure is time consuming and therefore not suitable for time evolution reconstructions. Taking a few sample points in time, one can however retrieve generic upper bounds for the uncertainty as a function of the phase-space contribution.

\subsection{Interpretation of momentum-space distribution}
\label{sec:ch6.3}

The reconstructed momentum-space distribution is examined in more detail to assess its plausibility. Since $E \gg E_c$, runaways should primarily be generated through the avalanche mechanism, causing their distribution function to decay exponentially with momentum \cite{Fulop2006}. For momentum-space contribution function values above a 10 percent of the maximum value, this behaviour is indeed found for the flux surface integrated reconstructed distribution from figure \ref{fig:10}a. Figure \ref{fig:12}a shows the distribution as a function of momentum for a pitch angle of 0.40$\,$rad at time $t = 1.357 \,$s. Above $p/{m_{\mathrm{e}}}c \sim 40$, $\log_{10}(f_{\mathrm{e}})$ decays linearly with $p$. The behaviour is characterized by some decay constant $C_{p}$, which from a theoretical point of view is expected to be a function of pitch angle:

\begin{equation}
    \sum_{\rho} f_{\mathrm{e}}(\rho,p,\theta_{\mathrm{p}})|_{\theta_{\mathrm{p}} = \theta_{\mathrm{p,0}}} = \mathrm{const} \cdot \exp \left[-C_{p}(\theta_{\mathrm{p,0}}) p\right],
\end{equation}

\noindent where $\theta_{\mathrm{p,0}}$ is the pitch angle under consideration. For various values of $\theta_{\mathrm{p,0}}$ the time evolution of the decay constant for $GJf_{\mathrm{e}}>0.1 \cdot \mathrm{max}(GJf_{\mathrm{e}})$, as obtained using a least-squares fit, is shown in figure \ref{fig:12}b. In general it is observed that the decay constant decreases over time. The equilibrium value of $C_p$ is $\theta_p$ dependent, the larger pitch decay constants being found at larger pitch angles. This is consistent with collisional kinetic theory and results from the balance between electric field acceleration and pitch angle scattering.

An explanation for the time dependency of $C_p$ is found by examining the time evolution of the flux surface integrated momentum-space profile. The main change in time is the increase in relative occupation of the high momentum part of phase-space. Figure \ref{fig:13}a depicts the evolution of the contour line at 5 percent of the maximum of the flux surface integrated momentum profile. The front moves to higher momenta as the runaways are accelerated, as indicated in figure \ref{fig:13}b. This momentum increase manifests itself in a decrease in $C_p$.

The rise in total recorded synchrotron radiation, as shown in figure \ref{fig:3}f, is attributed to an increase of the occupation of the high momentum, high pitch angle part of momentum-space over time. Reconstruction results indicate a population increase over the whole contributing region by a factor of 4.5 or more, which is consistent with the observed integrated intensity trend. Given the rather constant PMTX and MultiCam noise signals, see figure \ref{fig:3}e, it is not expected that the overall runaway current changes much after 1$\,$s. The main rise in synchrotron signal however occurs between 1 and 1.2$\,$s. This discrepancy is likely related to the time required for newly generated runaways to accelerate to the dominantly contributing energies.

The location of the contribution peak at a pitch angle of 0.53$\,$rad is harder to explain. As was pointed out by Hoppe et al. for a similar TCV discharge \cite{Hoppe2020}, this value strongly exceeds theoretical expectations. High-frequency kinetic instabilities were proposed as a pitch angle scattering mechanism which could give rise to these anomalously high values of $\theta_{\mathrm{p}}$ \cite{Liu2018}. For the discussed experiment no diagnostic capable of probing such perturbations, starting at frequencies of a several hundred MHz, is in place. An alternative pitch angle scattering mechanism is the magnetic field ripple \cite{Russo1991}, as will be discussed in the next section. At the time of writing no experiments have been performed to address either of these hypotheses.

\subsection{Interpretation of flux surface distribution}
\label{sec:ch6.4}

Until this point in the discussion, the flux surface dependence of the distribution has been disregarded. In the momentum-space region dominating the emission, the reconstruction indicates that the runaway profile peaks at the edge, as shown in figure \ref{fig:10}c. No statements are made on the radial distribution at low energies and pitch angles, as these are poorly constrained. The same is true for the distribution shape in the plasma centre. $GJf_{\mathrm{e}}$ contribution values indicate that conclusions about the runaway density profile can only be drawn for the machine minor radii $\rho>0.4$.

The rather surprising increase of the radial density profile between $\rho = 0.4$ and $\rho = 1$ could find its origin in the same mechanism as the anomalously high observed pitch angles. Possibly, the mechanism scattering the REs perpendicular to the field lines significantly gains in strength towards the edge of the plasma. 

A pitch angle scattering mechanism which lines up with this hypothesis is the magnetic field ripple. Resonant interaction can occur between runaways gyrating at their relativistically downshifted electron cyclotron frequency and the modulations in the toroidal field caused by the finite amount of toroidal field coils \cite{Russo1991}. The value of $\gamma$ for which the resonance occurs is roughly

\begin{equation}
    \label{eq: ripple}
    \gamma_{\mathrm{ripple}} = \frac{e B R}{m_{\mathrm{e}} c n_{\mathrm{res}} N_{\mathrm{coils}}},
\end{equation}

\noindent where $n_{\mathrm{res}}$ is the resonant mode number and $N_{\mathrm{coils}}$ is the number of toroidal field coils. For shot 64717, the $n_{\mathrm{res}} = 1-3$ resonances are expected to occur near normalized momenta of 46, 23 and 15 respectively. The interaction strength decreases with mode number, and increases moving closer towards the coils at larger minor radius \cite{Martin1999}. Runaways accelerating through the resonances are expected to undergo significant pitch angle scattering. Because of the increasing interaction strength moving towards the edge of the plasma, the momentum-space region dominating the synchrotron emission could thus realistically have the a shape depicted in figure \ref{fig:10}c. Note that the part of momentum-space contributing to the image has normalized momentum starting at roughly 40, which could indicate interaction of the runaways with the $n_{\mathrm{res}} = 1$ resonance. Testing of the ripple hypothesis would be aided by experiments in which the magnetic field strength is varied.

As pointed out by Hoppe et al \cite{Hoppe2020}, the high pitch angles could also result from an interaction with high frequency plasma instabilities. Strong experimental evidence of such events backed up by kinetic simulations has been reported for DIII-D \cite{Liu2018,Paz_Soldan_2019}. For the present experiments, this mechanism does not directly explain the increase in population towards the edge of the plasma at high pitch angles and momenta. A mechanism which could account for this profile is the build up of a runaway shell at the edge of the plasma during startup. As the loop voltage is initially peaked at the edge a local runaway shell could form. Due to its high conductivity it could act as a partial barrier for the electric field, sustaining an edge peaked profile for the runaways while allowing Ohmic current drive in the core.

\section{Summary and conclusions}
\label{sec:ch7}

This work elaborates on recent advances in multispectral camera analysis of runaway synchrotron radiation. 2D synchrotron profiles obtained through absolutely calibrated cameras with the same field-of-view but different central filter wavelength, provide both spatial and spectral resolution. It is found that:

\begin{itemize}
\item[] The spatial and spectral resolution of runaway synchrotron patterns offered by multispectral camera systems provide a complementary widening of the probed region of momentum-space.
\end{itemize}

Strengthened by the improved level of constraint on the runaway distribution, the tomographic SART method is used for the first time to provide a distribution reconstruction with no constraint on the momentum-space profile. The outlined approach relies on the synthetic diagnostic SOFT to provide a response function linking an arbitrary distribution in guiding-center coordinates to pixel intensities on a specified detector. Using a test distribution to generate a set of corresponding images through a response function, the performance of the approach is assessed. The three main conclusions are:

\begin{itemize}
    \item[i)] Provided an absolutely calibrated camera system the SART algorithm can be used to locally reconstruct the high momentum, high pitch angle part of the runaway phase-space distribution with an accuracy better than a factor of 2. No constraints on the shape of the momentum- or spatial-profile are required, and sharp gradients in the distribution can be dealt with.
    \item[ii)] The contribution of each phase-space element to reconstructed images is a robust measure for the level of trust to be placed in local reconstruction accuracy.
    \item[iii)] Increasing the number of spectral channels used as constraints for the reconstruction improves the accuracy of the distribution reconstruction.
\end{itemize}

Next, the algorithm is put to the test using multispectral synchrotron data from the MultiCam system in TCV tokamak. The observed patterns, with highly distinctive geometric features, are accurately reconstructed. It is found that the runaways dominating the emission have $p/{m_{\mathrm{e}}}c$ in excess of 40 and pitch angles above 0.30$\,$rad. The contribution peak is found at $(p/{m_{\mathrm{e}}}c,\theta_{\mathrm{p}}) = (60, 0.53 \,$rad), consistent with earlier estimates using a super-particle approach \cite{Hoppe2020}. At these high momenta and pitch angles the reconstructed profile increases above a normalized radius of $\rho > 0.4$ towards the edge. Note that this does not imply that the overall runaway current exhibits a hollow profile.

The physical plausibility of the TCV runaway reconstruction is addressed. It is found that:

\begin{itemize}
    \item[] Runaway distribution reconstructions from TCV multispectral sychrotron data show an exponential decay with increasing particle momentum, as is expected for an avalanche-dominated distribution in equilibrium. Time sequence reconstructions show a relaxation of the momentum distribution with a pitch angle dependency consistent with collisional kinetic theory.
\end{itemize}

\noindent Additional experiments are needed to understand the pitch angle and spatial distribution:

\begin{itemize}
    \item[] Pitch angles in the phase-space region contributing most to the synchrotron image are anomolously high. The toroidal field ripple is proposed as a candidate for causing pitch angle scattering. This picture is consistent with an increase in occupation of these high pitch angles moving from the core of the plasma towards the edge. An alternative explanation is found in high frequency plasma instabilities in combination with the formation of a runaway shell at the edge during startup, limiting electric field penetration towards the core of the plasma.
\end{itemize}

Further experiments are needed to address both hypotheses. These efforts will be strengthened by the enhanced runaway distribution details which the advances in multispectral imaging systems and the use of the SART method enable.


\ack

DIFFER is part of the institutes organisation of NWO. This work has been carried out within the framework of the EUROfusion Consortium and has received funding from the Euratom research and training programme 2014-2018 and 2019-2020 under grant agreement No 633053. The views and opinions expressed herein do not necessarily reflect those of the European Commission. 

\section*{References}
\bibliographystyle{unsrturl}
\bibliography{bibfile}

\begin{thebibliography}{10}

\bibitem{Hollmann2015}
E.M. Hollmann, P.B. Aleynikov, T.~F{\"{u}}l{\"{o}}p, D.A. Humphreys, V.A. Izzo,
  M.~Lehnen, V.E. Lukash, G.~Papp, G.~Pautasso, F.~Saint-Laurent, and J.A.
  Snipes.
\newblock Status of research toward the {ITER} disruption mitigation system.
\newblock {\em Physics of Plasmas}, 22(2):021802, 2015.
\newblock \href {https://doi.org/10.1063/1.4901251}
  {\path{doi:10.1063/1.4901251}}.

\bibitem{Lehnen2015}
M.~Lehnen et~al.
\newblock Disruptions in {ITER} and strategies for their control and
  mitigation.
\newblock {\em Journal of Nuclear Materials}, 463:39 -- 48, 2015.
\newblock \href {https://doi.org/10.1016/j.jnucmat.2014.10.075}
  {\path{doi:10.1016/j.jnucmat.2014.10.075}}.

\bibitem{Breizman2019}
B.N. Breizman, P.~Aleynikov, E.M. Hollmann, and M.~Lehnen.
\newblock Physics of runaway electrons in tokamaks.
\newblock {\em Nuclear Fusion}, 59(8):083001, 2019.
\newblock \href {https://doi.org/10.1088/1741-4326/ab1822}
  {\path{doi:10.1088/1741-4326/ab1822}}.

\bibitem{Hoppe2018_1}
M.~Hoppe, O.~Embr{\'{e}}us, C.~Paz-Soldan, R.A. Moyer, and
  T.~F{\"{u}}l{\"{o}}p.
\newblock Interpretation of runaway electron synchrotron and bremsstrahlung
  images.
\newblock {\em Nuclear Fusion}, 58(8):082001, 2018.
\newblock \href {https://doi.org/10.1088/1741-4326/aaae15}
  {\path{doi:10.1088/1741-4326/aaae15}}.

\bibitem{Finken1990}
K.H. Finken, J.G. Watkins, D.~Rusb\"{u}ldt, W.J. Corbett, K.H. Dippel, D.M.
  Goebel, and R.A. Moyer.
\newblock Observation of infrared synchrotron radiation from tokamak runaway
  electrons in {TEXTOR}.
\newblock {\em Nuclear Fusion}, 30(5):859--870, 1990.
\newblock \href {https://doi.org/10.1088/0029-5515/30/5/005}
  {\path{doi:10.1088/0029-5515/30/5/005}}.

\bibitem{Jaspers1994}
R.~Jaspers, N.J. Lopes~Cardozo, K.H. Finken, B.C. Schokker, G.~Mank, G.~Fuchs,
  and F.C. Sch\"uller.
\newblock Islands of runaway electrons in the {TEXTOR} tokamak and relation to
  transport in a stochastic field.
\newblock {\em Physical Review Letters}, 72:4093--4096, 1994.
\newblock \href {https://doi.org/10.1103/PhysRevLett.72.4093}
  {\path{doi:10.1103/PhysRevLett.72.4093}}.

\bibitem{Tinguely2018_1}
R.A. Tinguely, R.S. Granetz, M.~Hoppe, and O.~Embr{\'{e}}us.
\newblock Spatiotemporal evolution of runaway electrons from synchrotron images
  in {Alcator C-Mod}.
\newblock {\em Plasma Physics and Controlled Fusion}, 60(12):124001, 2018.
\newblock \href {https://doi.org/10.1088/1361-6587/aae6ba}
  {\path{doi:10.1088/1361-6587/aae6ba}}.

\bibitem{Tinguely2018_2}
R.A. Tinguely, R.S. Granetz, M.~Hoppe, and O.~Embr{\'{e}}us.
\newblock Measurements of runaway electron synchrotron spectra at high magnetic
  fields in {Alcator C-Mod}.
\newblock {\em Nuclear Fusion}, 58(7):076019, 2018.
\newblock \href {https://doi.org/10.1088/1741-4326/aac444}
  {\path{doi:10.1088/1741-4326/aac444}}.

\bibitem{Stahl2013}
A.~Stahl, M.~Landreman, G.~Papp, E.~Hollmann, and T.~F\"{u}l\"{o}p.
\newblock Synchrotron radiation from a runaway electron distribution in
  tokamaks.
\newblock {\em Physics of Plasmas}, 20(9):093302, 2013.
\newblock \href {https://doi.org/10.1063/1.4821823}
  {\path{doi:10.1063/1.4821823}}.

\bibitem{Shi2010}
Y.~Shi, J.~Fu, J.~Li, Y.~Yang, F.~Wang, Y.~Li, W.~Zhang, B.~Wan, and Z.~Chen.
\newblock Observation of runaway electron beams by visible color camera in the
  {Experimental Advanced Superconducting Tokamak}.
\newblock {\em Review of Scientific Instruments}, 81(3):033506, 2010.
\newblock \href {https://doi.org/10.1063/1.3340909}
  {\path{doi:10.1063/1.3340909}}.

\bibitem{Yu2013}
J.H. Yu, E.M. Hollmann, N.~Commaux, N.W. Eidietis, D.A. Humphreys, A.N. James,
  T.C. Jernigan, and R.A. Moyer.
\newblock Visible imaging and spectroscopy of disruption runaway electrons in
  {DIII-D}.
\newblock {\em Physics of Plasmas}, 20(4):042113, 2013.
\newblock \href {https://doi.org/10.1063/1.4801738}
  {\path{doi:10.1063/1.4801738}}.

\bibitem{Zhou2014}
R.J. Zhou, I.M. Pankratov, L.Q. Hu, M.~Xu, and J.H. Yang.
\newblock Synchrotron radiation spectra and synchrotron radiation spot shape of
  runaway electrons in {Experimental Advanced Superconducting Tokamak}.
\newblock {\em Physics of Plasmas}, 21(6):063302, 2014.
\newblock \href {https://doi.org/10.1063/1.4881469}
  {\path{doi:10.1063/1.4881469}}.

\bibitem{Hoppe2018_2}
M.~Hoppe, O.~Embr{\'{e}}us, R.A. Tinguely, R.S. Granetz, A.~Stahl, and
  T.~F\"{u}l\"{o}p.
\newblock {SOFT}: a synthetic synchrotron diagnostic for runaway electrons.
\newblock {\em {Nuclear Fusion}}, 58(2):026032, 2018.
\newblock \href {https://doi.org/10.1088/1741-4326/aa9abb}
  {\path{doi:10.1088/1741-4326/aa9abb}}.

\bibitem{Hoppe2020}
M.~Hoppe, G.~Papp, T.A. Wijkamp, J.~Decker, B.P. Duval, O.~Embreus,
  T.~F{\"{u}}l{\"{o}}p, U.~A. Sheikh, {TCV Team}, and {EUROfusion MST1 Team}.
\newblock Runaway electron synchrotron radiation in a vertically translated
  plasma.
\newblock {\em Nuclear Fusion}, 60(9):094002, 2020.
\newblock \href {https://doi.org/10.1088/1741-4326/aba371}
  {\path{doi:10.1088/1741-4326/aba371}}.

\bibitem{Hoppe2020_2}
M.~Hoppe, L.~Hesslow, O.~Embreus, L.~Unnerfelt, G.~Papp, I.~Pusztai,
  T.~F\"{u}l\"{o}p, O.~Lexell, T.~Lunt, E.~Macusova, P.J. McCarthy,
  G.~Pautasso, G.I. Pokol, G.~Por, P.~Svensson, {the ASDEX Upgrade team}, and
  {the EUROfusion MST1 team}.
\newblock Spatiotemporal analysis of the runaway distribution function from
  synchrotron images in an {ASDEX} {Upgrade} disruption.
\newblock {\em Under consideration in Journal of Plasma Physics}, 2020.
\newblock URL: \url{https://arxiv.org/abs/2005.14593}.

\bibitem{Perek2019}
A.~Perek, W.A.J. Vijvers, Y.~Andrebe, I.G.J. Classen, B.P. Duval, C.~Galperti,
  J.R. Harrison, B.L. Linehan, T.~Ravensbergen, K.~Verhaegh, and M.R. de~Baar.
\newblock {MANTIS: A real-time quantitative multispectral imaging system for
  fusion plasmas}.
\newblock {\em {Review of Scientific Instruments}}, 90(12):123514, 2019.
\newblock \href {https://doi.org/10.1063/1.5115569}
  {\path{doi:10.1063/1.5115569}}.

\bibitem{Perek2020}
A.~Perek, B.L. Linehan, M.~Wensing, K.~Verhaegh, I.G.J. Classen, B.P. Duval,
  O.~F\'{e}vrier, H.~Reimerdes, C.~Theiler, T.~A. Wijkamp, M.R. {de Baar}, {the
  EUROfusion MST1 team}, and {the TCV team}.
\newblock {Measurement of the 2D emission profiles of the bulk and impurity
  ions in the TCV divertor}.
\newblock {\em Submitted to Journal of Nuclear Materials and Energy}, 2020.
\newblock \href {https://doi.org/10.13140/RG.2.2.21477.42720}
  {\path{doi:10.13140/RG.2.2.21477.42720}}.

\bibitem{Calcam}
S.~Silburn, J.~Harrison, M.~Smithies, A.~Wynn, T.~Farley, and J.~Cavalier.
\newblock Calcam (version 2.5.0).
\newblock \url{https://github.com/euratom-software/calcam/tree/v2.2.0}, 2020.
\newblock \href {https://doi.org/10.5281/zenodo.3956834}
  {\path{doi:10.5281/zenodo.3956834}}.

\bibitem{Hawke2017}
J.~Hawke, Y.~Andrebe, R.~Bertizzolo, P.~Blanchard, R.~Chavan, J.~Decker, B.P.
  Duval, P.~Lavanchy, X.~Llobet, B.~Marl{\'{e}}taz, P.~Marmillod, G.~Pochon,
  and M.~Toussaint.
\newblock Improving spatial and spectral resolution of {TCV} thomson
  scattering.
\newblock {\em {Journal of Instrumentation}}, 12(12):C12005--C12005, 2017.
\newblock \href {https://doi.org/10.1088/1748-0221/12/12/c12005}
  {\path{doi:10.1088/1748-0221/12/12/c12005}}.

\bibitem{Connor1975}
J.W. Connor and R.J. Hastie.
\newblock Relativistic limitations on runaway electrons.
\newblock {\em Nuclear Fusion}, 15(3):415--424, 1975.
\newblock \href {https://doi.org/10.1088/0029-5515/15/3/007}
  {\path{doi:10.1088/0029-5515/15/3/007}}.

\bibitem{Moret2015}
J.M. Moret, B.P. Duval, H.B. Le, S.~Coda, F.~Felici, and H.~Reimerdes.
\newblock Tokamak equilibrium reconstruction code {LIUQE} and its real time
  implementation.
\newblock {\em Fusion Engineering and Design}, 91:1 -- 15, 2015.
\newblock \href {https://doi.org/10.1016/j.fusengdes.2014.09.019}
  {\path{doi:10.1016/j.fusengdes.2014.09.019}}.

\bibitem{Guan2010}
X.~Guan, H.~Qin, and N.J. Fisch.
\newblock Phase-space dynamics of runaway electrons in tokamaks.
\newblock {\em Physics of Plasmas}, 17(9):092502, 2010.
\newblock \href {https://doi.org/10.1063/1.3476268}
  {\path{doi:10.1063/1.3476268}}.

\bibitem{Andersen1984}
A.H. Andersen and A.C. Kak.
\newblock Simultaneous algebraic reconstruction technique ({SART}): A superior
  implementation of the art algorithm.
\newblock {\em Ultrasonic Imaging}, 6(1):81 -- 94, 1984.
\newblock \href {https://doi.org/10.1016/0161-7346(84)90008-7}
  {\path{doi:10.1016/0161-7346(84)90008-7}}.

\bibitem{Carr2018}
M.~Carr, A.~Meakins, M.~Bernert, P.~David, C.~Giroud, J.~Harrison,
  S.~Henderson, B.~Lipschultz, and F.~Reimold.
\newblock Description of complex viewing geometries of fusion tomography
  diagnostics by ray-tracing.
\newblock {\em Review of Scientific Instruments}, 89(8):083506, 2018.
\newblock \href {https://doi.org/10.1063/1.5031087}
  {\path{doi:10.1063/1.5031087}}.

\bibitem{Fulop2006}
T.~F\"{u}l\"{o}p, G.~Pokol, P.~Helander, and M.~Lisak.
\newblock Destabilization of magnetosonic-whistler waves by a relativistic
  runaway beam.
\newblock {\em Physics of Plasmas}, 13(6):062506, 2006.
\newblock \href {https://doi.org/10.1063/1.2208327}
  {\path{doi:10.1063/1.2208327}}.

\bibitem{Guo2018}
Z.~Guo, C.J. McDevitt, and X.~Tang.
\newblock Control of runaway electron energy using externally injected whistler
  waves.
\newblock {\em Physics of Plasmas}, 25(3):032504, 2018.
\newblock \href {https://doi.org/10.1063/1.5019381}
  {\path{doi:10.1063/1.5019381}}.

\bibitem{Liu2018}
C.~Liu, E.~Hirvijoki, G.~Fu, D.P. Brennan, A.~Bhattacharjee, and C.~Paz-Soldan.
\newblock Role of kinetic instability in runaway-electron avalanches and
  elevated critical electric fields.
\newblock {\em Physical Review Letters}, 120:265001, 2018.
\newblock \href {https://doi.org/10.1103/PhysRevLett.120.265001}
  {\path{doi:10.1103/PhysRevLett.120.265001}}.

\bibitem{Russo1991}
A.J. Russo.
\newblock Effect of ripple on runaway electrons in tokamaks.
\newblock {\em Nuclear Fusion}, 31(1):117--126, 1991.
\newblock \href {https://doi.org/10.1088/0029-5515/31/1/011}
  {\path{doi:10.1088/0029-5515/31/1/011}}.

\bibitem{Martin1999}
J.R. Mart\'{i}n-Sol\'{i}s, B.~Esposito, R.~S\'{a}nchez, and J.D. Alvarez.
\newblock Energy limits on runaway electrons in tokamak plasmas.
\newblock {\em Physics of Plasmas}, 6(1):238--252, 1999.
\newblock \href {https://doi.org/10.1063/1.873276}
  {\path{doi:10.1063/1.873276}}.

\bibitem{Paz_Soldan_2019}
Paz-Soldan C., N.W. Eidietis, E.M. Hollmann, P.~Aleynikov, L.~Carbajal, W.W.
  Heidbrink, M.~Hoppe, C.~Liu, A.~Lvovskiy, D.~Shiraki, D.~Spong, D.P. Brennan,
  C.M. Cooper, D.~del Castillo-Negrete, X.~Du, O.~Embreus, T.~Fulop,
  J.~Herfindal, R.~Moyer, P.~Parks, and K.E. Thome.
\newblock Recent {DIII-D} advances in runaway electron measurement and model
  validation.
\newblock {\em Nuclear Fusion}, 59(6):066025, 2019.
\newblock \href {https://doi.org/10.1088/1741-4326/ab1769}
  {\path{doi:10.1088/1741-4326/ab1769}}.

\end{thebibliography}

\end{document}